\documentclass{article}
\usepackage[utf8]{inputenc}
\usepackage[english]{babel}
\usepackage{graphicx}
\usepackage{subcaption}
\usepackage{amssymb}
\usepackage{dirtytalk}
\usepackage{listings}
\usepackage{csquotes}
\usepackage{amsmath}
\usepackage{amsthm}
\usepackage{multirow}
\usepackage{algpseudocode} 
\usepackage{algorithm} 
\usepackage{xcolor}
\usepackage{dirtytalk}
\usepackage{authblk}
\usepackage{lineno}

\theoremstyle{definition}

\usepackage[margin=0.9in]{geometry}

\usepackage{comment} 

\usepackage[
backend=biber,
style=numeric,
maxnames=19,
sorting=none,
citestyle=numeric
]{biblatex}
\title{Predicting Lung Cancer's Metastats' Locations Using Bioclinical Model}

\author[1*]{Teddy Lazebnik}
\author[2]{Svetlana Bunimovich-Mendrazitsky}

\affil[1]{Department of Cancer Biology, Cancer Institute, University College London, London, UK}
\affil[2]{Department of Mathematics, Ariel University, Israel}
\affil[*]{Corresponding author: lazebnik.teddy@gmail.com}

\date{}

\begin{document}

\maketitle
\begin{abstract}
Lung cancer is a leading cause of cancer-related deaths worldwide. The spread of the disease from its primary site to other parts of the lungs, known as metastasis, significantly impacts the course of treatment. Early identification of metastatic lesions is crucial for prompt and effective treatment, but conventional imaging techniques have limitations in detecting small metastases. In this study, we develop a bioclinical model for predicting the spatial spread of lung cancer's metastasis using a three-dimensional computed tomography (CT) scan. We used a three-layer biological model of cancer spread to predict locations with a high probability of metastasis colonization. We validated the bioclinical model on real-world data from 10 patients, showing promising \(74\%\) accuracy in the metastasis location prediction. Our study highlights the potential of the combination of biophysical and ML models to advance the way that lung cancer is diagnosed and treated, by providing a more comprehensive understanding of the spread of the disease and informing treatment decisions. 

\end{abstract}

\linenumbers
\nolinenumbers
\section{Background}
Treating cancer is a critical challenge in modern medicine, as it affects millions of people worldwide and can have devastating effects on both patient health and quality of life. Lung cancer, in particular, is one of the most prevalent and deadly forms of cancer worldwide \cite{Torre2016}. One of the major challenges in treating lung cancer is the development of metastases, which are secondary tumors that develop from cells that have spread from the primary tumor site to other parts of the organ (or even the body) \cite{intro_1,intro_2,intro_3}. The location of these metastases can greatly affect a patient's prognosis and the effectiveness of treatment \cite{intro_4}.

In particular, for lung cancer diagnosis, healthcare professionals commonly use a \textit{computed tomography} (CT) and Positron Emission Tomography CT (PET-CT) to diagnose the patient's clinical condition \cite{intro_5}. This non-invasive imaging modality allows for the simultaneous acquisition of functional and anatomic information, providing detailed insights into the metabolic activity of the tumor and its relationship to surrounding tissue. Additionally, this analysis provides the anatomy and structure of the lung and surrounding tissue, which can be used to determine the size and location of the tumor. 
Based on this data, clinicians look for the primary cancer tumor's as well as metastasis' properties to determine the course of treatment \cite{intro_6,intro_7}. 

Recently, to partially automate the process of parsing the required clinical data from CT imagery, researchers have used computer vision and ML algorithms in general and to detect lung cancer, in particular \cite{rw_1,rw_3,liza_2, Tandon22, Lakshmanaprabu}. These algorithms can (semi-)automatically detect and classify lung nodules, reducing the dependence on human interpretation and improving the consistency of diagnoses. In addition, biophysical models gathering popularity in the clinical domain as these getting better at predicting clinical outcomes over time \cite{models_good_1,models_good_2,models_good_3,models_good_4,models_good_5}. For instance, \cite{intro_example_ode} constructs a mathematical model that integrates let-7 and miR-9 expression into a signaling pathway to generate an \textit{in silico} model for the process of epithelial-mesenchymal transition. The authors validate their model using \textit{in vitro} data collected by testing the effects of EGFR inhibition on downstream MYC, miR-9, and let-7a expression.  \cite{intro_example_ode_2} propose a multicomponent mathematical model for simulating lung cancer growth as well as radiotherapy treatment for lung cancer, showing promising predictive accuracy compared to a relatively small set of \textit{in vivo} data points. 

Unfortunately, current PET-CT technology is lacking the ability to clearly capture small metastasis, usually less than 2 mm in diameter due to the mix of cancer and healthy cells together with the noise occurring during the sampling process \cite{intro_11}. Missing small-size metastasis can lead to choosing the sub-optimal course of treatment and result in clinically catastrophic results. In this work, we propose a bio-physical model to predict from PET-CT imagery the locations of small-size metastasis that the current PET-CT methods are missing. Data-driven models are increasingly being utilized in the field of oncology \cite{kidney_injury, review_ML}. By utilizing patient-specific biological and clinical data, these models aim to provide a more comprehensive representation of the disease and its progression, thereby enabling more targeted and effective treatment strategies \cite{intro_8,intro_9}. The use of these models in the treatment of lung cancer holds great promise, and this study aims to evaluate their ability to predict the location of lung cancer metastases, with the ultimate goal of improving patient outcomes \cite{intro_10}.

The novelty of this work lies in the combination of three types of bio-physical models associated with cancer settlement, flow in the bloodstream, and growth inside healthy tissue. We tested the proposed model on the historical data of 10 lung cancer patients with metastasis, obtaining a 74\% accuracy in the prediction of the metastasis locations. 

The remainder of the paper is organized as follows. Section \ref{section:model} outlines the proposed bioclinical model, divided into the biophysical model, initial condition construction from the 3D CT image, and the metastasis probability heatmap generation process. Section \ref{sec:results} presents the performance of the proposed model on real-world clinical data. Finally, section \ref{sec:disccusion} discusses the obtained results with their clinical applications and proposes possible future work.

\section{Model definition}
\label{section:model}

\begin{figure}
    \centering
    \includegraphics[width=0.99\textwidth]{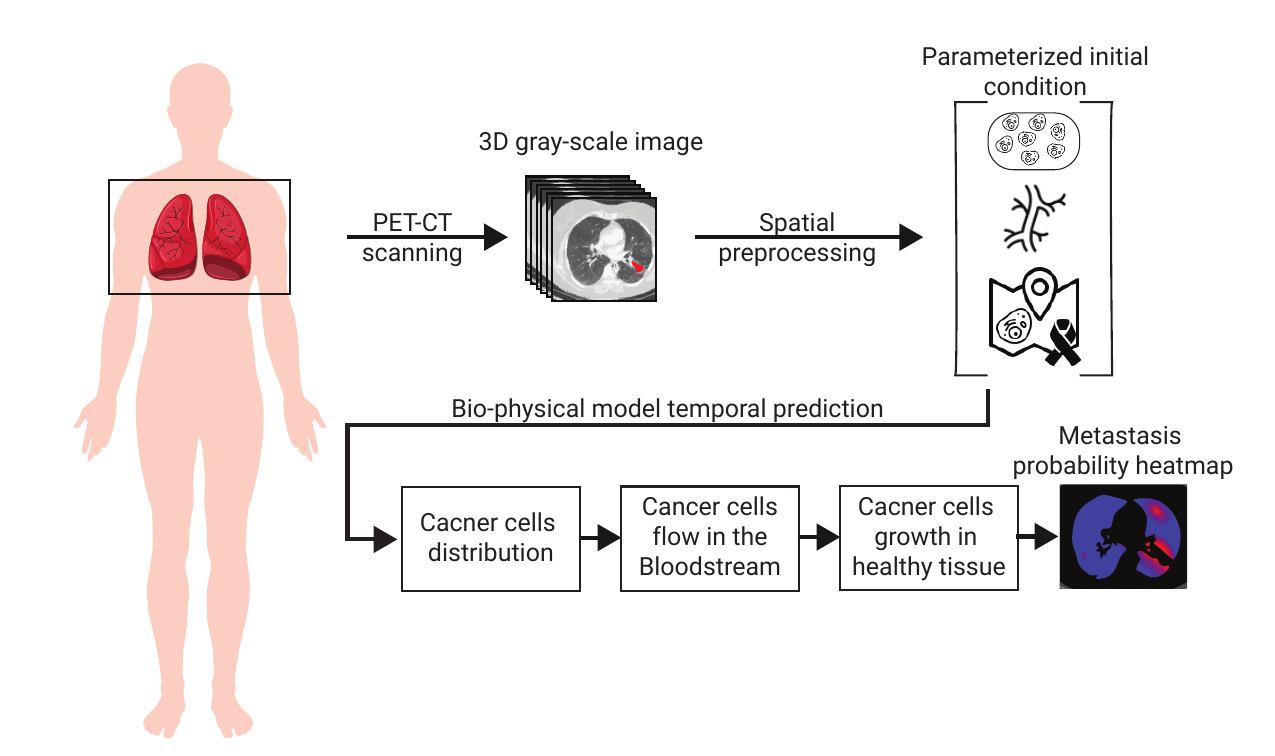}
    \caption{A schematic view of the model's components and the interactions between them.}
    \label{fig:scheme}
\end{figure}

The spread dynamics of metastasis originating in the primary tumor can be associated with three main biological processes: 1) the flow of cancer cells in the bloodstream; 2) the settlement of cancer cells in the tissue; and 3) the spatial spread of cancer polyps \cite{md_start_1,md_start_2,md_start_3}. All these biological processes are spatio-temporal in nature and therefore influenced by the spatial distributions of blood vessels, healthy tissue, and the original cancer cells. To capture these settings, we use a chest PET-CT image, obtaining a three-dimensional (3D) gray-scale image \((I \in \mathbb{R}^{x \times y \times z})\) where \(x, y, z\) are the CT image's dimensions. Using \(I\), and by simulating the biological and clinical occurring \textit{in vivo}, one can predict the locations of cancer's metastasis, if these exist. A schematic view of the model's components and the interactions between them is summarised in Fig. \ref{fig:scheme}. Namely, the algorithm can be divided into three main components. First, biophysical modeling is responsible for predicting the ability of cancer cells to colonize different parts of the lungs over time. Second, the computer vision algorithm accepts the 3D CT image and produces the parameterized initial condition construction for the biophysical model. Finally, an algorithm that utilizes the biophysical model to generate the metastasis heatmap.  

\subsection{Biophysical modeling}
\label{section:bio_model}
To capture the biological processes occurring \textit{in vivo} in a patient's lungs in the context of lung cancer migration, we define a model, \(M\). Formally, \(M\) is a function \(M: \mathbb{G} \times \mathbb{R}^3 \times \{0, 1\}^{x \times y \times z} \times \mathbb{R}^3  \rightarrow \mathbb{R}^{+}\) such that \(M(G, p[l], I, \tau) \rightarrow c\), where \(G \in \mathbb{G}\) is the flow graph, \(p[l] \in \mathbb{R}^3\) is the location of the primary tumor in the image \(I\), and \(I \in \{0, 1\}^{x \times y \times z}\) is a 3D binary image with \(1\) for locations where cancer cells can settle and \(0\) otherwise, \(\tau \in \mathbb{R}^3\) is the location of interest to evaluate in the image \(I\), and \(c \in \mathbb{N}\) is the total number of cancer cells predicted to settle in location \(\tau\). 

Formally, the model operates as follows. First, we find the blood vessel \(b_s \in B_s\) that is closest to the primary tumor location (\(p[l]\)) by computing the distance between the primary tumor location and each blood vessel \(\forall b_s \in B_s\) and taking the minimal value. Afterward, in the same manner, the blood vessel closest to \(\bar{x}\). These two locations are marked by \(\bar{S} \in b_s^S\) and \(\bar{T} \in b_s^T\), respectively. Afterward, we find all the paths in the flow graph \((G)\) shorter than a length \(\nu\) such that starts at \(b_s^S\) and ends at \(b_s^T\) using the breadth-first search (BFS) algorithm \cite{bfs}.

It is known that the cancer cell population's size is reducing exponentially over time \cite{pk_sample,pk_sample_2}. Hence, if the absolute value of cancer cells in the cardiovascular is less than a pre-defined threshold \(\xi \in \mathbb{N}\), the population size is set to be zero and by that defines the value of \(\nu\) and the stop condition for this step in the model. 

Hence, we proposed a prediction algorithm \(A\) that gets as an input the location of a primary tumor \(\tau_l\) and its size \(\tau_s\), the blood vessels, and the connections between them as a graph \(G = (B, E)\), and a binary 3D tensor of healthy tissue cancer metastasis can colonize and returns a 3D tensor indicating the normalized probability that metastasis would take occur in each segment of the lungs. Namely, one can define algorithm \(A\) as follows: \(A: \mathbb{R}^{x \cdot y \cdot z} \times \mathbb{R}^3 \times \mathbb{R} \rightarrow \mathbb{R}^{x \cdot y \cdot z}\) such that \(A(I, \tau_l, \tau_s) \rightarrow P(I)\), where \(I \in \mathbb{R}^{x \cdot y \cdot z}\) is the CT image, \(\tau_l \in \mathbb{R}^3\) is the location of the prime tumor in \(I\), \(\tau_s \in \mathbb{R}\) is the size of the primary tumor, and \(P(I) \subset \mathbb{R}^{x \cdot y \cdot z}\) is a distribution function allocating a probability of metastasis occurrence for each location in \(I\).

\(A\) operates as follows, starting from the primary tumor, the tumor grows according to the model proposed by \cite{cancer_growth} until it reaches a blood vessel. At this point, cancer cells are assumed to drop from the tumor into the bloodstream at some rate \(d \in \mathbb{R}^+\). The cancer cells are floating in the bloodstream according to the model proposed by \cite{teddy_nano_2}. In a probabilistic manner, cancer cells leave the bloodstream toward the tissues around the blood vessels and try to grow into a metastasis. This colonization process is governed by the model proposed by \cite{model_colonize}. 
 
\subsection{Initial condition construction}
\label{section:intial_condition}
In order to utilize the proposed bio-physical with an inputted PET-CT 3D image, one first needs to process the image to the required parameterized format that can be used as an initial condition for the model. Namely, it is required to extract the location and size of the primary tumor, the blood vessels in the geometry, and the locations of healthy tissue the cancer cells can colonize.  

The first task can be achieved with relatively high accuracy using the method proposed by \cite{cancer_from_ct}, extracting its location (\(\tau_l \in \mathbb{R}^3\)) and size \(\tau_s \in \mathbb{R}\) in \(I\). 

Next, the blood vessels are defined using a graph  \(G = (B, E)\) such that \(E \subset B \times B\). The blood vessels graph, \(G\), is obtained as follows. First, a search radius \(R \in \mathbb{R} > 0 \) is initialized manually by the user to be \(R_0\). The blood vessels approximated by a cylinder geometry from a 3D CT image are obtained using the algorithm proposed by \cite{blood_vessels_from_images}. We denoted the set of blood vessels to be \(B\) where \(b \in B: b := (c, h, r, o_{xy}, o_{xz})\) such that \(c \in \mathbb{R}^3\) is the blood vessel's center of mass, \(h \in \mathbb{R}\) is the height of the blood vessel, \(r \in \mathbb{R}\) is the average radius of the blood vessel, \(o_{xy} \in [0, 2\pi]\) is the orientation in the \(xy\) axis, and   \(o_{xz} \in [0, \pi]\) is the orientation in the \(xz\) axis. For convenience, we treat \(B\) as a list that is sorted according to the \(h\) parameter, which can be obtained in \(O(|B|log(|B|))\) operations using Quicksort algorithm \cite{quicksort}. For each \(b \in B\), we compute \(sl := c - h/2 [cos (o_{xy}), cos (\pi/4 - o_{xy}), cos (o_{xz})] \) and \(el := c + h/2 [cos (o_{xy}), cos (\pi/4 - o_{xy}), cos (o_{xz})]\) which stands for the start and end locations of a cylinder blood vessel, respectively. To contract \(G\), we first initialize it such that \(B\) are the nodes and \(E\) is an empty edge set. Now, in an interactive manner, we add edges to \(E\) with the following logic: each blood vessel, \(b \in B\) is checked if its 
start, \(sl\), is inside a 3D sphere defined by the search radius \(R\) with either the beginning or end location of another blood vessel. If so, we add it to the edge set, \(E\), such that its end is connected to the closest blood vessel \(b \in B\) in terms of \(||b^i_{sl} - b^j_{el}||\) where \(b^i, b_j \in B \wedge b^i \neq b_j\). If no edge is added, the search radius, \(R\), is increased according to the formula \(R \leftarrow R + \delta_R\) such that \(\delta_R \in \mathbb{R}^+\) is a pre-define hyperparameter. Once \(G\) becomes a connected graph, the process holds. This is due to the fact that it is biologically known that the cardiovascular system is connected and as such, once this criterion is met, it is assumed the graph is properly constructed. Finally. we compute the maximum (in the manner of radius) spanning tree of the graph \(G\) using the method proposed by Al Mamun and Rajasekaran \cite{spin_tree}. 

Finally, the 3D image is first divided into 2D images. Then, the location of the healthy tissue is obtained by using a threshold-adoptive Canny algorithm \cite{canny} to first find the edges between the inner part of the lungs and the outer one, alongside other edges in the image that are noise. In order to remove the unwanted edges, we used Hough Transform \cite{hough_trans} to detect two ellipsoids from the edges that approximate the outer divide edges. Afterward, the edges connected to the ones obtained from the Hough Transform's ellipsoids, are added until a close polygon is obtained. The inner side of these two shapes is defined to be the healthy tissue area. After the two shapes are obtained for each 2D image, we reconstruct the entire 3D shape using the Laplacian smoothing method \cite{smooth_three_d}.

\subsection{Metastasis heatmap generation}
\label{section:heatmap_generation}
Hypothetically, one could solve the heatmap generation as a regression task, solving the model numerically with a given initial condition at time \(t_0 \in \mathbb{R}\) and extracting the state of the model at any desired time, \(t_f > t_0\). However, as shown in \cite{teddy_nano_2}, computing such a dynamical system numerically is both computationally expensive and unstable. The main issue lies in the graph-based Navier-Stokes equations \cite{ns_reduce} one needs to solve. As such, to tackle this challenge, we generate the metastasis heatmap as follows. An 3D binary (\(I\)) is uniformly sampled with a grid \(\Omega\) of size \(\alpha_x, \alpha_y, \alpha_z \in \mathbb{N}\) for the \(x, y, z\) axes, respectively. The probability of metastasis occurrence in each point in \(\Omega\) is obtained by solving the biophysical model (see Section \ref{section:bio_model}) with the initial condition (see Section \ref{section:intial_condition}) with one modification. The blood flow component is computed for only the path in \(G\) from the initial tumor location and the tested point in the grid. Once all points in the grid are tested and the amount of cancer cells is computed for a pre-defined stop time \(T \in \mathbb{R}^+\), the values are normalized using the \(L_1\) metric to define probability.

\section{Results}
\label{sec:results}
To evaluate the performance of the proposed bioclinical model, we used samples from 10 patients. All patients are diagnosed with metastases in the lungs in the Sheba Hospital (Israel) between 2019 and 2022. The metastasis instances have been manually tagged by a clinician, indicating the location in the lung and the size, alongside a manual tagging of the prime tumor's location and size. For these patients, a CT image before the detection of metastasis and afterward are obtained, without treatment in between, allowing to validate of the metastasis occurring locations without outside influence. 

Table \ref{table:results} outlines the result of this computation, divided into the Hard and Soft classifier scores. Overall, the bioclinical model shows 74\% accuracy, on average, in predicting the location of lung metastasis from lung cancer with a 1.9\% standard deviation. This indicates that the model receives stable levels of performance over the sampled population. The scores describe the accuracy metric of a Soft-classifier between two 3D images \(I_1 \in \{0, 1\}^{x \cdot y \cdot z}\) and \(I_2 \in \mathbb{R}^{x \cdot y \cdot z}\) where \(I_1\) is a binary image that indicates if a cell in the image is part of a cancer polyp or not and \(I_2\) is the prediction of the proposed model (see Section \ref{section:heatmap_generation}):

\begin{equation}
    \begin{array}{l}
         d_s(I_1, I_2) := 1 - \frac{\sum_{1 \leq i \leq x} \sum_{1 \leq j \leq y} \sum_{1 \leq k \leq z} (I_1[x,y,z] - I_2[x,y,z])^2}{x \cdot y \cdot z}
    \end{array}.
    \label{eq:performance}
\end{equation}

In addition, we formally define the Hard-classifier score to be \(d_h: \{0, 1\}^{x \cdot y \cdot z} \times \{0, 1\}^{x \cdot y \cdot z} \times  \mathbb{R} \rightarrow \mathbb{R} \) such that \(d_h(I_1, I_2, \zeta) := d_s(I_1, I_2|_\zeta)\) where \(I_2|_\zeta\) is obtained by performing a threshold \(\zeta\) on the predicted image \(I_2\). I.e. each value \(\alpha \in I_2\) is replaced with \(1\) if \(\alpha > \zeta\) and \(0\) otherwise). 

\begin{table}[!ht]
\centering
\begin{tabular}{|c|c|c|}
\hline
\multicolumn{1}{|l|}{\textbf{Patient}} & \multicolumn{1}{l|}{\textbf{Hard-classifier score}} & \multicolumn{1}{l|}{\textbf{Soft-classifier score}} \\ \hline
\textbf{1} & 0.782 & 0.803 \\ \hline
\textbf{2} & 0.830 & 0.838 \\ \hline
\textbf{3} & 0.689 & 0.706 \\ \hline
\textbf{4} & 0.656 & 0.663 \\ \hline
\textbf{5} & 0.735 & 0.742 \\ \hline
\textbf{6} & 0.835 & 0.851 \\ \hline
\textbf{7} & 0.721 & 0.733 \\ \hline
\textbf{8} & 0.674 & 0.691 \\ \hline
\textbf{9} & 0.701 & 0.714 \\ \hline
\textbf{10} & 0.690 & 0.708 \\ \hline
\textbf{\(MEAN \pm STD\)} & \(0.7313 \pm 0.0192\) & \(0.7449 \pm 0.0193\) \\ \hline
\end{tabular}
\caption{The proposed model's performance, divided into patients.}
\label{table:results}
\end{table}

\section{Discussion}
\label{sec:disccusion}
The spread of lung cancer from its primary site to other parts of the lungs is a critical aspect of lung cancer progression and is associated with decreased survival rates \cite{dis_1,des_2}. Early identification of metastatic lesions is crucial for prompt and effective treatment, as it can significantly impact patient outcomes \cite{des_3}. However, conventional imaging techniques such as computed tomography (CT) scans have limitations in detecting small metastases \cite{gili_met}. In this study, we proposed a personalized biophysical-based bioclinical mathematical model that accepts a patient's 3D CT image and produces a 3D heatmap of the probability a metastasis would develop in each region of the lungs. 

\begin{figure}
    \centering
    \includegraphics[width=0.5\textwidth]{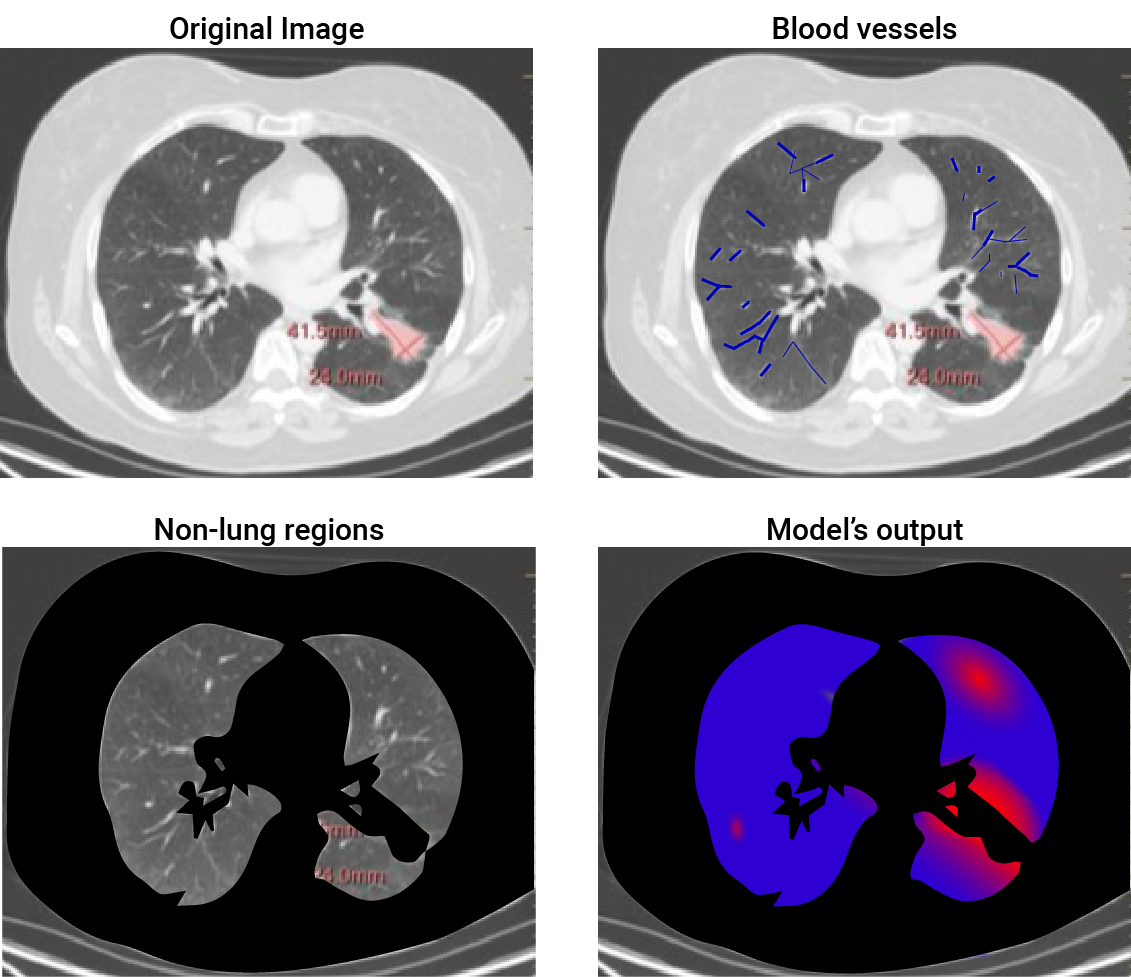}
    \caption{A 2D slice of patient \#2.}
    \label{fig:paitent_view}
\end{figure}

We tested our bioclinical model on real-world clinical data and validated it with clinician domain experts to get a baseline. As seen in Table \ref{table:results}, the proposed bioclinical model provides around 74\% accuracy in predicting the location of metastasis. This means, that on average, three out of four predictions of the model marked locations with metastasis in the CT that would not be marked otherwise. However, these results do not provide a lower-boundary accuracy measurement of the model's performance as locations marked by the algorithm with metastasis that was without are either a wrong prediction or that metastasis was not found during the validation phase and would be created given more time. For example, let us examine a 2D slice of one of the patients in the sample. Fig. \ref{fig:paitent_view} shows the original CT image slice (with the manual mark of the primary tumor), the blood vessels marked in blue, the healthy tissue locations, and the model's heatmap prediction - from left to right and top to bottom. One can notice that the model predicts metastasis for the right-top corner of the lungs in this z-axis value. However, metastasis is not found there. This is because either the model is mistaken or not enough time passed from the first and second CT scans in order to allow metastasis to grow into a detected size. 

One of the key advantages of the bioclinical model is the fact it is based on biophysical modeling of the cancer spread which allows one to integrate diverse data sources and provides a more comprehensive understanding of the disease compared to purely data-driven methods. Theoretically, one could aim to develop a data-driven solution to this time-series task using ML or deep learning methods. However, to do so, an extensive data-gathering process of multiple CT scans of lung cancer would be required, which would be extremely expensive, logistically complex, and ethically questionable. On the other hand, the proposed approach utilizes previous biological, physical, and clinical knowledge, resulting in a need for just a relatively small dataset, used entirely to validate and evaluate the performance of the proposed model.  

The results of this study demonstrate the potential of mathematical models in predicting the location of lung cancer metastases. Our findings show that the bioclinical model is capable of providing personalized predictions of metastatic spread with decent levels of accuracy. Hence, this model has the potential to inform more targeted and effective treatment strategies for lung cancer patients, ultimately improving patient outcomes. As such, this study provides a step forward in the development of more personalized and effective treatments for lung cancer. The use of bio-clinical mathematical models to predict the location of metastases holds great promise for improving patient outcomes and should be a focus of future research efforts in this field.

However, our bioclinical model can (and should) be further improved. First, the currently used biomathematical model of cancer colonization on healthy tissue does not take into consideration the spatial properties of the cancer cells. Second, cancer cells continue to mutate over time which causes a wide range of outcomes and can significantly alter the course of metastasis formation. Third, taking into consideration additional clinical data about the patient such as gender, background diseases, smoking, and others, it will be possible to obtain more accurate the location of cancer metastases \cite{liza_1}. In addition, this study was limited by its sample size and the heterogeneity of the patient population. Thus, additional large-scale, multicentered trials are necessary to further test the accuracy and clinical utility of the proposed model, to obtain more statistically representative results.

\section*{Declarations}

\subsection*{Ethical Statement}
None

\subsection*{Data availability}
The data used in this research is available by a formal request from the authors. Regulations according to the Israeli Ministry of Healthcare might be applied. 

\subsection*{Conflict of Interests}
None.

\subsection*{Author Contributions}
Teddy Lazebnik: Conceptualization, Methodology, Software, Validation, Formal analysis, Investigation, Resources, Data Curation, Writing - Original Draft, Writing - Review \& Editing, Visualization, Project administration. 
Svetlana Bunimovich-Mendrazitsky: Data Curation, Writing - Review \& Editing, Funding acquisition.

\subsection*{Acknowledgements}
The authors wish to thank Stephen Raskin for providing the data and tagging used in this study.

\printbibliography

@article{kidney_injury,
    author = {Lazebnik, T. and Bahouth, Z. and Bunimovich-Mendrazitsky, S. and  Halachmi, S.},
    title = {Predicting acute kidney injury following open partial nephrectomy treatment using SAT-pruned explainable machine learning model},
    year = {2022},
    volume = {22},
    pages = {133-140},
    number = {1},
    journal = {BMC Medical Informatics and Decision Making},
}

@article{review_ML,
    author = {Yu, X. and Ji, Y. and Huang, M. and Feng, Z.},
    title = {Machine learning for acute kidney injury: Changing the traditional disease prediction mode},
    year = {2023},
    volume = {10:1050255},
    pages = {1-21},
    journal = {Front Med},
}

@article{gili_met,
    author = {Hochman, G. and Shacham-Shmueli, E. and Raskin, S.P. and Rosenbaum, S. and Bunimovich-Mendrazitsky, S.},
    title = {Metastasis Initiation Precedes Detection of Primary Cancer-Analysis of Metastasis Growth in vivo in a Colorectal Cancer Test Case},
    year = {2020},
    volume = {33391005},
    pages = {1-9},
    journal = {Front Physiol.},
}

@article{Tandon22,
    author = {Tandon, R. and Agrawal, S. and Chang, A. and Band, S.S.},
    title = {Hybrid Deep Learning Model for Detection and Classification of Lung Carcinoma Using Chest Radiographs},
    year = {2022},
    volume = {894920},
    pages = {1-13},
    journal = {Front Public Health.},
}

@article{Lakshmanaprabu,
    author = {Lakshmanaprabu, S.K. and Mohanty, S.N. and  Shankar, K. and  Arunkumar, N. and  Ramirez, G.},
    title = {Optimal deep learning model for classification of lung cancer on CT images},
    year = {2019},
    volume = {92},
    pages = {374-382},
    journal = {Future Generat. Comput. Syst}
}

@inproceedings{spin_tree,
  author={Al Mamun, A. and Rajasekaran, S.},
  booktitle={2016 IEEE Symposium on Computers and Communication (ISCC)}, 
  title={An efficient Minimum Spanning Tree algorithm}, 
  year={2016},
  pages={1047-1052},
}

@article{bfs,
  title =	 {A new distributed breadth-first-search algorithm},
  author =	 {Zhu, Y. Cheung, Y-T.},
  journal =	 {Information Processing Letters},
  volume =	 {25},
  number = {5},
  year =	 {1987},
  pages =	 {329-333},
}

@article {pk_sample,
	author = {Ben-Akiva, E. and Meyer, R. A. and Yu, H. and Smith, J. T. and Pardoll, D. M. and Green, J. J.},
	title = {Biomimetic anisotropic polymeric nanoparticles coated with red blood cell membranes for enhanced circulation and toxin removal},
	volume = {6},
	number = {16},
	year = {2020},
	journal = {Science Advances}
}

@article{pk_sample_2,
    author = {Lee, M. J-E. and Veiseh, O. and Bhattarai, N. and Sun, C. and Hansen, S. J. and Ditzler, S. and Knoblaugh, S. and Lee, D. and Ellenbogen, R. and Zhang, M. and Oslon, J. M.},
    year = {2010},
    title = {Rapid Pharmacokinetic and Biodistribution Studies Using Cholorotoxin-Conjugated Iron Oxide Nanoparticles: A Novel Non-Radioactive Method},
    journal = {Plos one},
    volume = {5},
    number = {3},
    pages = {e9536}
}

@INPROCEEDINGS{cancer_from_ct,
  author={Miah, B. A. and Yousuf, M. A.},
  booktitle={2015 International Conference on Electrical Engineering and Information Communication Technology (ICEEICT)}, 
  title={Detection of lung cancer from CT image using image processing and neural network}, 
  year={2015},
  pages={1-6},
}

@Inbook{Torre2016,
author="Torre, L. A.
and Siegel, R. L.
and Jemal, A. ",
editor="Ahmad, A.
and Gadgeel, S.",
title="Lung Cancer Statistics",
bookTitle="Lung Cancer and Personalized Medicine: Current Knowledge and Therapies",
year="2016",
}

@article{intro_1,
    author = {Yuan, M. and Huang, L-L. and Chen, J-H. and Wu, J. and Xu, Q.},
    year = {2019},
    title = {The emerging treatment landscape of targeted therapy in non-small-cell lung cancer},
    journal = {Signal Transduction and Targeted Therapy},
    volume = {4},
    pages = {61}
}

@article{intro_2,
    author = {Reck, M. and Popat, S. and Reinmuth, N. and  De Ruysscher, D. and Kerr, K. M. and Peters, S.},
    year = {2014},
    title = {Metastatic non-small-cell lung cancer (NSCLC): ESMO Clinical Practice Guidelines for diagnosis, treatment and follow-up},
    journal = {Annals of Oncology},
    volume = {25},
    pages = {III27-III39}
}

@article{intro_3,
    author = {Wang, M. and Herbst, R. S. and Boshoff, C.},
    year = {2021},
    title = {Toward personalized treatment approaches for non-small-cell lung cancer},
    journal = {Nature Medicine},
    volume = {27},
    pages = {1345–1356}
}

@article{intro_4,
    author = {Popper, H. H. },
    year = {2016},
    title = {Progression and metastasis of lung cancer},
    journal = {Cancer Metastasis Review},
    volume = {35},
    pages = {75–91}
}

@article{intro_5,
title = {PET/CT imaging in different types of lung cancer: An overview},
journal = {European Journal of Radiology},
volume = {81},
number = {5},
pages = {988-1001},
year = {2012},
author = {Ambrosini, V. and Nicolini, S. and Caroli, P. and Nanni, C. and Massaro, A.  and Cristina, M. M. and Rubello, D. and Fanti, S. }
}

@article{intro_6,
author = {He, Y-Q. and Gong, H-L. and Deng, Y-F. and Li, W-M.},
title ={Diagnostic efficacy of PET and PET/CT for recurrent lung cancer: a meta-analysis},
journal = {Acta Radiologica},
volume = {55},
number = {3},
pages = {309-317},
year = {2014}
}

@article{intro_7,
title={PET/CT imaging in lung cancer: indications and findings},
volume={41},  
year = {2015},
volume = {41},
number = {3},
journal={Jornal Brasileiro de Pneumologia},  
author={Hochhegger, B. and Alves, G. R. T. and Irion, K. L. and Fritscher, C. C. and Fritscher, L. G. and Concatto, N. H. and Marchiori, E.}
}

@article{intro_8,
title = {Personalized Therapy for Lung Cancer},
journal = {Chest},
volume = {146},
number = {6},
pages = {1649-1657},
year = {2014},
author = {Moreira, A. L. and Eng, J.},
}

@article{intro_9,
title = {Personalized therapy for lung cancer: striking a moving target},
journal = {JCI Insight},
volume = {3},
number = {15},
pages = {e120858},
year = {2018},
author = {Pakkala, S. and Ramalingam, S. S.},
}

@article{intro_10,
author = {Kim, S. M. and Lee, H. and Min, B-H. and Kim, J. J. and An, J. Y. and Choi, M-G. and Bae, J. M. and Kim, S. and Sohn, T. S. and Lee, J. H.},
title = {A prediction model for lymph node metastasis in early-stage gastric cancer: Toward tailored lymphadenectomy},
journal = {Journal of Surgical Oncology},
volume = {120},
number = {4},
pages = {670-675},
year = {2019}
}

@article{intro_11,
author = {Wang, P. and DeNunzio, A. and Okunieff, P. and O'Dell, W. G.},
title = {Lung metastases detection in CT images using 3D template matching},
journal = {Medical Physics},
volume = {34},
number = {3},
pages = {915-922},
year = {2007}
}

@article{rw_1,
title = {Optimal deep learning model for classification of lung cancer on CT images},
journal = {Future Generation Computer Systems},
volume = {92},
pages = {374-382},
year = {2019},
author = {Lakshmanaprabu, S.K. and Sachi, N. M. and Shankar, K. and Arunkumar, N. and Gustavo, R.},
}

@article{rw_3,
author = {Guo, Z. and Xu, L. and Si, Y. and Razmjooy, N.},
title = {Novel computer-aided lung cancer detection based on convolutional neural network-based and feature-based classifiers using metaheuristics},
journal = {International Journal of Imaging Systems and Technology},
volume = {31},
number = {4},
pages = {1954-1969},
year = {2021}
}

@INPROCEEDINGS{blood_vessels_from_images,
    author={Kirbas, C. and Quek, F. K. H.},
    booktitle={Third IEEE Symposium on Bioinformatics and Bioengineering, 2003. Proceedings.},   
    title={Vessel extraction techniques and algorithms: a survey},
    year={2003},
    pages={238-245}
}

@article{quicksort,
  title={Quicksort},
  author={Hoare, C. A. R.},
  journal={The Computer Journal},
  volume={5},
  number={1},
  pages={10--16},
  year={1962}
}

@article{canny,
  title={A computational approach to edge detection},
  author={Canny, J.},
  journal={IEEE Transactions on pattern analysis and machine intelligence},
  number={6},
  pages={679--698},
  year={1986}
}

@article{hough_trans,
  title={Use of the Hough transformation to detect lines and curves in pictures},
  author={Duda, R. O. and Hart, P. E.},
  journal={Communications of the ACM},
  volume={15},
  number={1},
  pages={11--15},
  year={1972},
}

@InProceedings{smooth_three_d,
author="Volodine, T.
and Vanderstraeten, D.
and Roose, D.",
editor="Kim, M-S.
and Shimada, K.",
title="Smoothing of Meshes and Point Clouds Using Weighted Geometry-Aware Bases",
booktitle="Geometric Modeling and Processing - GMP 2006",
year="2006",
pages="687--693",
}

@article{teddy_nano_2,
  title={Generic Purpose Pharmacokinetics-Pharmacodynamics Mathematical Model For Nanomedicine Targeted Drug Delivery: Mouse Model},
  author={Lazebnik, T. and Weitman, H. and Kaminka, G. A.},
  journal={bioRxiv},
  year={2022}
}

@Article{models_good_1,
AUTHOR = {Lazebnik, T.},
TITLE = {Cell-Level Spatio-Temporal Model for a Bacillus Calmette-Guérin-Based Immunotherapy Treatment Protocol of Superficial Bladder Cancer},
JOURNAL = {Cells},
VOLUME = {11},
YEAR = {2022},
NUMBER = {15},
ARTICLE-NUMBER = {2372},
}

@article{models_good_2,
    author={Umar, M. and Sabir, Z. and Amin, F. Guirao, J. L. G. and Raja, M. A. Z. },
    title={Stochastic numerical technique for solving HIV infection model of CD4+ T cells},
    journal={The European Physical Journal Plus}, 
    year={2020},
    volume = {135},
    pages = {403}
}

@article{models_good_3,
    author = {De Pillis, L. G. and  Gu, W. and  Radunskaya, A. E.},
    journal = {Journal of Theoretical Biology},
    title = {Mixed immunotherapy and chemotherapy of tumors:
modeling, applications and biological interpretations},
    year = {2006},
    volume = {238},
    pages = {841–862},
}

@article{models_good_4,
    author = {Eikenberry, S. and Thalhauser, C. and Kuang, Y.},
    title = {Tumor-immune interaction, surgical treatment, and cancer recurrence in a mathematical model of melanoma},
    journal={Plos Computational Biology},
    year = {2009},
    pages = {e1000362}
}

@article{models_good_5,
title = {A mathematical model for pancreatic cancer growth and treatments},
journal = {Journal of Theoretical Biology},
volume = {351},
pages = {74-82},
year = {2014},
author = {Louzoun, Y. and Xue, C. and Lesinski, G. B. and Friedman, A.},
}

@article{intro_example_ode,
    author = {Kang, H-W. AND Crawford, M. AND Fabbri, M. AND Nuovo, G. AND Garofalo, M. AND Nana-Sinkam, S. P. AND Friedman, A.},
    journal = {PLOS ONE},
    title = {A Mathematical Model for MicroRNA in Lung Cancer},
    year = {2013},
    volume = {8},
    pages = {1-19},
    number = {1},

}

@article{intro_example_ode_2,
    author = {Hong, W-S. and Wang, S-Q. and Zhang, G-Q.},
    journal = {Computational and Mathematical Methods in Medicine},
    title = {Lung Cancer Radiotherapy: Simulation and Analysis Based on a Multicomponent Mathematical Model},
    year = {2021},
    volume = {2021},
    pages = {6640051},

}

@article{model_colonize,
title = {Modeling the inhibition of breast cancer growth by GM-CSF},
journal = {Journal of Theoretical Biology},
volume = {303},
pages = {141-151},
author = {Szomolay, B. and Eubank, T. D. and Roberts, R. D. and Marsh, C. B. and Friedman, A.},
}

@article{cancer_growth,
    author = {Newton, P. K. AND Mason, j. AND Bethel, K. AND Bazhenova, L. A. AND Nieva, J. AND Kuhn, P.},
    journal = {PLOS ONE},
    title = {A Stochastic Markov Chain Model to Describe Lung Cancer Growth and Metastasis},
    year = {2012},
    volume = {7},
    pages = {1-18},
}

@article{md_start_1,
    author = {Larsen, J. E. and Minna, J. D. },
    journal = {Clinics in Chest Medicine},
    title = {Molecular Biology of Lung Cancer: Clinical Implications},
    year = {2011},
    volume = {32},
    number = {4},
    pages = {703-740},
}

@article{md_start_2,
title = {Emerging Biological Principles of Metastasis},
journal = {Cell},
volume = {168},
number = {4},
pages = {670-691},
year = {2017},
author = {Lambert, A. W. and Pattabiraman, D. R. and Weinberg, R. A.}
}

@article{md_start_3,
    author = {Fidler, I. J.},
    title = "{Tumor Heterogeneity and the Biology of Cancer Invasion and Metastasis1}",
    journal = {Cancer Research},
    volume = {38},
    number = {9},
    pages = {2651-2660},
    year = {1978}
}

@article{ns_reduce,
  author    = {Canic, S. and Mikelic, A. and Tambaca, J.}, 
  title     = {A two-dimensional effective model describing fluid–structure interaction in blood flow: analysis, simulation and experimental validation},
  journal = {Comptes Rendus Mecanique},
  year      = {2005},
  volume    = {333},
  number = {12},
  pages      = {867-883}
}

@article{dis_1,
title = {The role of the tumor-microenvironment in lung cancer-metastasis and its relationship to potential therapeutic targets},
journal = {Cancer Treatment Reviews},
volume = {40},
number = {4},
pages = {558-566},
year = {2014},
author = {Wood, S. L. and Pernemalm, M. and Crosbie, P. A. and Whetton. A. D.},
}

@article{des_2,
title = {A Japanese Lung Cancer Registry Study: Prognosis of 13,010 Resected Lung Cancers},
journal = {Journal of Thoracic Oncology},
volume = {3},
number = {1},
pages = {46-52},
year = {2008},
author = {Asamura, H. and Goya, T. and Koshiishi, Y. and Sohara, Y. and Eguchi, K. and Mori, M. and Nakanishi, Y. and Tsuchiya, R. and Shimokata, K. and Inoue, H. and Nukiwa, T. and Miyaoka, E.},
}

@article{des_3,
title = {Detection of cancer metastasis: past, present and future},
journal = {Clin Exp Metastasis},
volume = {39},
pages = {21-28},
year = {2022},
author = {Alix-Panabieres, C. and Magliocco, A. and  Cortes-Hernandez, L. E. and Eslami-S, Z. and Franklin, D. and Messina, J. L.},
}

@article{liza_1,
title = {Mathematical Modeling of BCG-based Bladder Cancer Treatment Using Socio-Demographics},
journal = {arXiv},
year = {2023},
author = {Savchenko, E. and Rosenfeld, A. and Bunimovich-Mendrazitsky, S.},
}

@article{liza_2,
title = {Computer aided functional style identification and correction in modern {Russian} texts},
journal = {Journal of Data, Information and Management },
year = {2022},
pages = {25-32},
volume = {4},
author = {Savchenko, E. and Lazebnik, T.},
}

\end{document}